\documentclass[%
 reprint,
 amsmath,amssymb,
 aps,
prl,
]{revtex4-2}

\usepackage{graphicx}
\usepackage{dcolumn}
\usepackage{bm}
\usepackage[colorlinks=true, allcolors=blue]{hyperref}
\usepackage{mathrsfs}%
\usepackage[title]{appendix}%
\usepackage{xcolor}%
\usepackage{textcomp}%
\usepackage{booktabs}%
\usepackage{algorithm}%
\usepackage{algorithmicx}%
\usepackage{algpseudocode}%
\usepackage{listings}%
\usepackage{float}
\usepackage{siunitx}

\begin{document}

\title{First Electron Acceleration in a Tunable-Velocity Laser Wakefield}

\author{Aaron Liberman}
\altaffiliation{These authors contributed equally \\ Corresponding Emails: \\ aaronrafael.liberman@weizmann.ac.il,\\ victor.malka@weizmann.ac.il}
\author{Anton Golovanov}%
\altaffiliation{These authors contributed equally \\ Corresponding Emails: \\ aaronrafael.liberman@weizmann.ac.il,\\ victor.malka@weizmann.ac.il}
\author{Slava Smartsev}%
\altaffiliation{These authors contributed equally \\ Corresponding Emails: \\ aaronrafael.liberman@weizmann.ac.il,\\ victor.malka@weizmann.ac.il}
\author{Anda-Maria Talposi}%
\altaffiliation{These authors contributed equally \\ Corresponding Emails: \\ aaronrafael.liberman@weizmann.ac.il,\\ victor.malka@weizmann.ac.il}
\author{Sheroy Tata}%
\author{Victor Malka}%
\affiliation{%
 Department of Physics of Complex Systems, Weizmann Institute of Science, Rehovot 7610001, Israel
}%

\date{\today}

\begin{abstract}
We present the first experimental confirmation that a laser-wakefield accelerator produced by a flying focus pulse is able to maintain the coherent structures necessary to accelerate electrons to relativistic energies. Through a combination of spatio-temporal near-field shaping of the beam and focusing with an axiparabola---a long-focal-depth mirror that produces a quasi-Bessel beam---the propagation velocity of the wakefield is tuned to control the maximum electron energy achievable. The experimental data are supported by advanced optical and particle-in-cell simulations and are aligned with a simplified analytical model. Together, the results significantly strengthen the case for the flying-focus wakefield as a strategy for mitigating dephasing in laser-wakefield acceleration. 

\end{abstract}

\maketitle


Since the seminal work of Tajima and Dawson \cite{Tajima_PRL_1979}, laser-wakefield accelerators (LWFAs) have been the subject of great interest in the scientific community for their ability to accelerate electrons with acceleration gradients orders of magnitude greater than those found in conventional accelerators \cite{Joshi_Nature_1984,Modena_Nature_1995}. LWFAs have demonstrated the ability to accelerate high quality, mono-energetic electron bunches \cite{Faure_Nature_2004,Geddes_Nature_2004,Mangles_Nature_2004} with electron energies of up to 10 GeV \cite{Picksley_PRL_2024}. They have also shown promise in applications ranging from next-generation light sources \cite{Wang_Nature_2021,Labat_NaturePhotonics_2022,LaBerge_NaturePhotonics_2024}, to novel cancer therapy treatments \cite{Glinec_Med_Phys_2006,Guo_NatureCommunications_2025}, and to the probing of strong-field QED \cite{Mirzaie_NaturePhotonics_2024,LUXE_EurPhysJ_2024}. However, in order to reach the parameter space needed for many applications, the maximum achievable electron energy must be further increased. One of the main limitations preventing the achievement of ever-higher electron energies is the dephasing limit, or the velocity mismatch between the trapped electrons and the wakefield which prematurely ends the acceleration process \cite{Joshi_Nature_1984,Esarey_ReviewOfModernPhysics_2009}. 

A number of proposals have been explored for mitigating the effects of dephasing. Among these are rephasing with a density upramp \cite{Sprangle_PRE_2001,Guillaume_PRL_2015,Gustafsson_ScientificReports_2024} and multi-staged LWFAs \cite{Leemans_PhysicsToday_2009}. The most successful solution to date has been to lower the plasma density in the LWFA \cite{Picksley_PRL_2024,Leemans_NaturePhysics_2006,Gonsalves_PRL_2019,Miao_PRX_2022}, thus reducing the mismatch between the group velocity of light in the plasma and the velocity of the electrons. While this method has successfully yielded electrons with energies of up to 10 GeV \cite{Picksley_PRL_2024, Rockafellow_PoP_2025}, it results in a lower acceleration gradient \cite{Tajima_PRL_1979} that requires guiding the laser pulse over long distances to prevent laser diffraction \cite{Esarey_ReviewOfModernPhysics_2009,Leemans_NaturePhysics_2006}. These two challenges become progressively more significant as the target electron energy increases, with the acceleration gradient becoming ever weaker and the length over which the pulse needs to be guided growing nonlinearly \cite{Esarey_ReviewOfModernPhysics_2009}. 

\begin{figure*} [t!]
   \begin{center}
    \includegraphics[width=\linewidth]{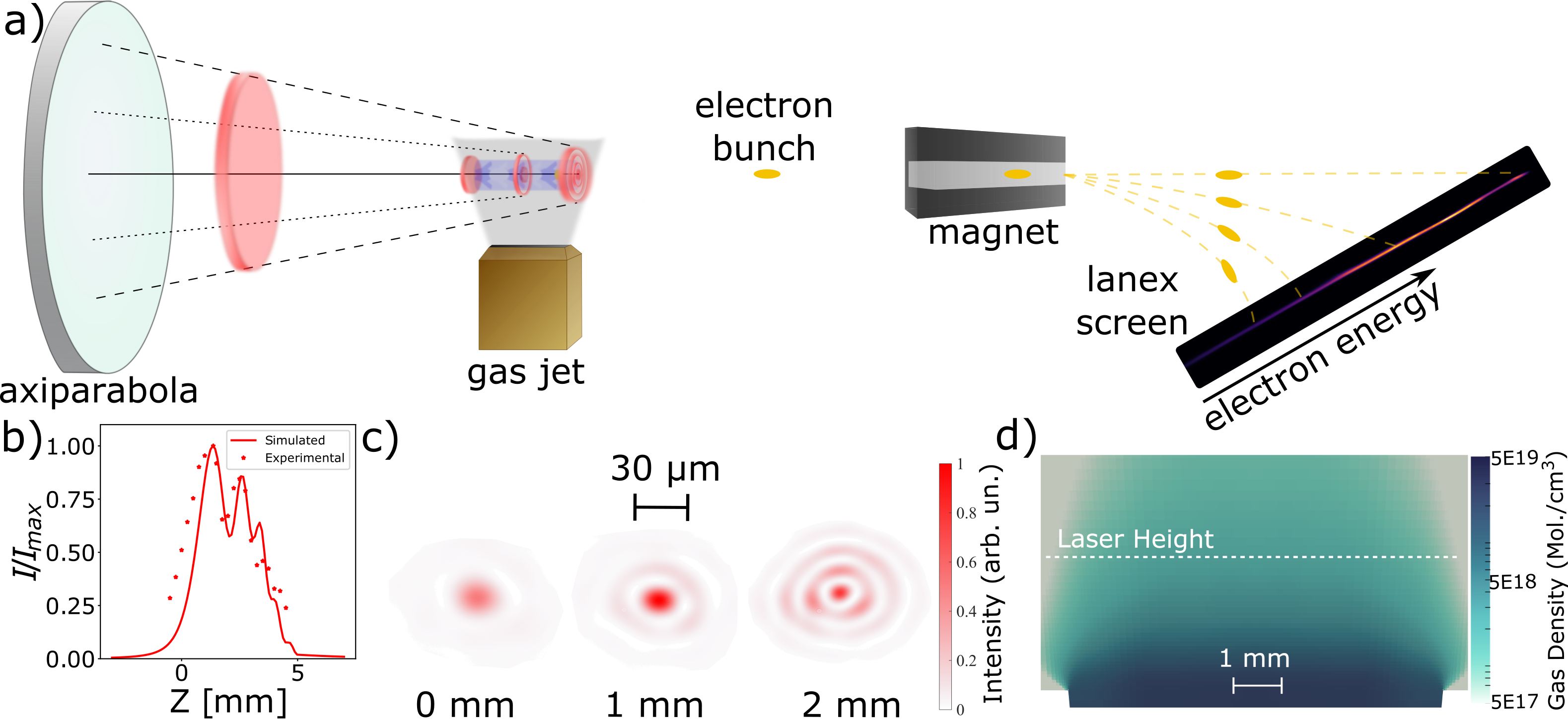}
    \end{center}
    \caption[]{ \label{fig:setup} (a) Schematic representation of the electron acceleration experiment. A laser pulse (red disk) is focused by the axiparabola (turquoise cylinder) onto a gas jet (gray emanating from gold jet), creating a wakefield (blue column) and accelerating an electron bunch (yellow dot). The laser pulse is also shown at focus over the gas jet demonstrating the development of the Bessel rings along the focal depth. After the gas jet, the electrons travel through a magnet, which gives an energy dependent trajectory to the electrons, which then impinge on a Lanex scintillator screen.  A sample Lanex image with an electron spectrum is shown. (b) Normalized focal spot intensity over the focal depth, in vacuum. Solid line shows simulated value while markers show experimentally measured points. (c) Selected 2D focal spots images at different points along the focal depth. (d) Ansys Fluent simulation of gas density from the nozzle. Dotted white line shows laser height.}
\end{figure*} 

A promising solution for dephasing is to use structured light to modify the on-axis propagation velocity of intensity, thus matching the propagation of the laser driver with the velocity of the trapped electrons \cite{Sainte-Marie_Optica_2017,Froula_NaturePhotonics_2018,Debus_PRX_2019,Caizergues_NaturePhotonics_2020,Palastro_PRL_2020}. This modification is achieved by inducing a ``flying focus'', a focus extended over several Rayleigh lengths, in which different rays focus to different points along the optical axis and the timing between these rays is tunable. Several schemes have been explored for achieving the flying focus. These include colliding two tilted laser pulses \cite{Debus_PRX_2019} and a combination of longitudinal chromatism and group-delay dispersion \cite{Sainte-Marie_Optica_2017,Froula_NaturePhotonics_2018}. However, these techniques are experimentally less practical for the high laser intensities necessary for LWFA. 

The flying-focus implementation that is most practical for LWFAs relies on a combination of spatio-temporal pulse shaping and focusing with a specialized optical element that produces a quasi-Bessel beam \cite{Caizergues_NaturePhotonics_2020,Palastro_PRL_2020}. This optic, known as the axiparabola \cite{Smartsev_OpticsLetters_2019,Oubrerie_JoO_2022}, induces both a focusing term and a controlled spherical aberration in the beam. This controlled aberration results in an extended focal depth \cite{Smartsev_OpticsLetters_2019} and changes the timing with which intensity focuses on the optical axis \cite{Caizergues_NaturePhotonics_2020,Palastro_PRL_2020}. This timing can be further modified by spatio-temporally shaping the pulse prior to focusing to add a radially dependent pulse-front delay \cite{Caizergues_NaturePhotonics_2020,Oubrerie_JoO_2022,Ambat_OpticsExpress_2023}. Simulations have shown that this combination can accelerate electrons with energies of over 100 GeV in mere meters of acceleration \cite{Caizergues_NaturePhotonics_2020,Palastro_PRL_2020}. The ability to tune the on-axis propagation velocity of intensity has been experimentally demonstrated \cite{Liberman_OL_2024,Pigeon_OE_2024}. Recent experimental results have revealed novel insights about the structure of the wakefield generated by such light pulses \cite{Liberman_NatureCommunications_2025}.

Here we present the first electrons accelerated in a wakefield generated by a laser pulse structured by an axiparabola and pulse-front curvature (PFC), a radially dependent pulse-front delay that depends quadratically on the radius. We show that the peak energy achieved in the acceleration depends on the PFC induced in the beam, and, therefore, on the wake velocity. The electron spectra and the energy-dependence on PFC are reproduced by particle-in-cell (PIC) simulations and supported by an analytical model. The successful acceleration of electrons with this structured wakefield serves as a critical proof-of-concept experiment that shows this novel wakefield is able to maintain the coherent structures necessary for accelerating relativistic electrons. A simulated snapshot of the wakefield directly shows the connection between the PFC in the beam and the propagation velocity. Thus, this is the first experimental confirmation of the effect of the propagation velocity of the wakefield on electron energy, suggesting the ability to partially mitigate dephasing effects in LWFA. 

The experiment was conducted using the HIGGINS 100 TW laser system at the Weizmann Institute of Science \cite{Kroupp_MRE_2022}. For the experiment, the Ti:Sapphire-based laser provided \SI{1.5}{J}, \SI{27}{fs} laser pulses with a \SI{30}{nm} bandwidth centered around \SI{800}{nm}. The \SI{50}{mm} diameter, quasi-top-hat beam was focused by a 10 degree off-axis axiparabola with a nominal focal length, $f_0$, of \SI{480}{mm} (f/9.6) and a focal depth, $\delta$, of \SI{5}{mm}. The axiparabola was designed so that the change in the focus as a function of radius, $r$, follows the expression: $ f(r) = f_0 + \delta (r/R)^2$ \cite{Oubrerie_JoO_2022}, where $R$ is the full aperture of the beam. 

\begin{figure*} [t!]
   \begin{center}
    \includegraphics[width=\linewidth]{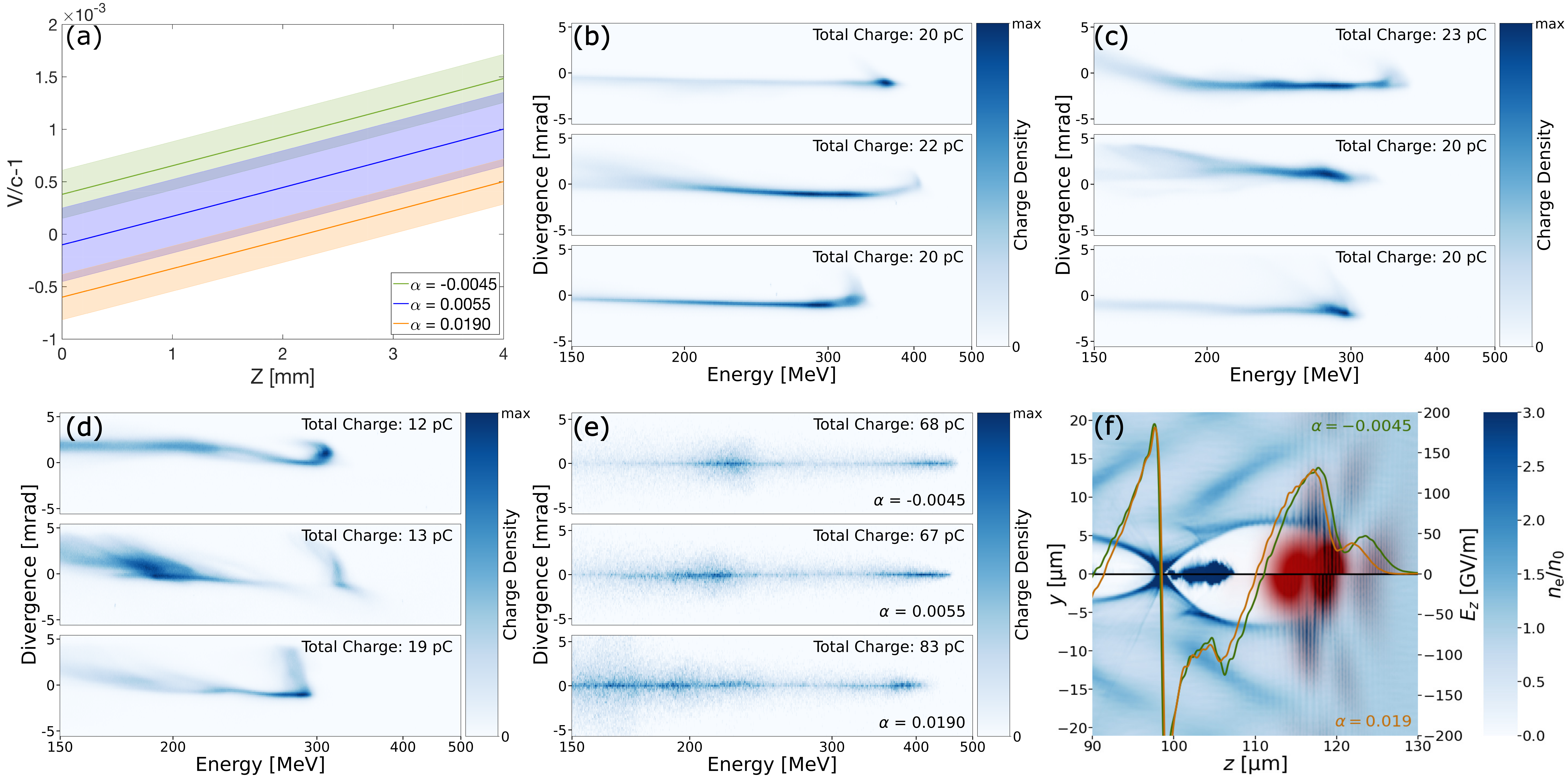}
    \end{center}
    \caption[]{ \label{fig:fig2} (a) Measured velocity  of intensity peak propagation in vacuum along the optical axis for the axiparabola focused beam. Shown for the $\alpha = -0.0045$ (green), $\alpha = 0.0055$ (blue), and $\alpha = 0.0190$ (orange) cases. The shaded area corresponds to the measurement error. (b--d) Three selected Lanex images for the $\alpha = -0.0045$ (b), $\alpha = 0.0055$ (c), and $\alpha = 0.0190$ (d) cases. The colorbar gives charge density, the x-axis provides energy information, and the y-axis gives the divergence. Total charge above 150 MeV is provided for each shot. (e) Simulated Lanex images for the $\alpha = -0.0045$ (top), $\alpha = 0.0055$ (middle), $\alpha = 0.0190$ (bottom) cases. (f) Simulated wakefield for $\alpha = -0.0045$ (top) and $\alpha = 0.0190$ (bottom) cases. Blue colorbar shows relative electron density distribution $n_e/n_0$ and red color shows the intensity of the axiparabola laser field. Line plots show the longitudinal electric field $E_z$ for the $\alpha = -0.0045$ (green) and $\alpha = 0.0190$ (orange) cases.}
\end{figure*}

Figure \ref{fig:setup} (a) shows a schematic of the experimental setup, in which the beam was focused by the axiparabola onto a 0.5 mm wide and 7 mm long supersonic slit nozzle with a \SI{500}{\um}$\times$ \SI{500}{\um} throat, generating a plasma with an electron density of around \SI{4e18}{cm^{-3}} at the laser height of $3$ mm from the nozzle exit. To facilitate ionization injection, a gas mixture of 97\% helium and 3\% nitrogen was used. The resultant electron bunch was then passed through a magnet and characterized on a Lanex screen. More details of the nozzle and Lanex can be found in Appendix A. Figure \ref{fig:setup} (b) shows the evolution of the normalized intensity over the course of the focal depth, both from the simulation (solid red line) and from measurements (red markers). Figure \ref{fig:setup} (c) shows three experimental 2D focal spot images at different points along the focal depth, illustrating the emergence of the Bessel-ring structure. The plasma density was determined by measuring the gas density of the nozzle using an interferometer and was confirmed with simulations conducted using the Ansys Fluent fluid simulation software. Figure \ref{fig:setup} (d) shows a 2D cut of the Fluent simulated gas density. As the gas is almost pure helium, the resulting plasma electron density is approximately twice the neutral gas density.

The PFC was measured using far-field beamlet cross-correlation \cite{Smartsev_JoO_2022}, which measures STCs by interfering two beamlets in the far-field, scanning the delay between them, and using inverse Fourier transform spectroscopy. To manipulate the PFC, a custom doublet lens was inserted into a beam expansion telescope stage of the laser system, before the grating compressor. The doublet modifies the PFC of the beam, with the magnitude of the modification depending on the beam size incident on the doublet \cite{Kabacinski_2021}. Therefore, moving the doublet inside of the beam expander allows for the tuning of the PFC of the beam \cite{Smartsev_JoO_2022, Liberman_OL_2024,Kabacinski_2021}. 

To simulate the electron acceleration process, a beam with parameters similar to the experimental beam was reflected from the axiparabola surface and propagated to the focal plane using the Axiprop code \cite{Andriyash_Axiprop,Oubrerie_JoO_2022}. The simulated on-axis intensity profile can be seen in figure \ref{fig:setup} (b). An additional PFC phase correction was applied to the beam before the reflection. The obtained laser field was then input into particle-in-cell (PIC) simulations with FBPIC \cite{Lehe_ComPhysCom_2016}, a quasi-3D spectral code that utilizes azimuthal mode decomposition, assuming linear polarization in the $x$ direction (p-polarization). The gas profile obtained from Fluent was used. The gas was initialized as neutral to account for diffraction at larger radii where the gas might be partially ionized. The Lanex was simulated by propagating the beam from PIC simulations through a model of the spectrometer magnet and projecting its density onto a planar screen. Further details of the simulation can be found in Appendix B. 

Electrons were accelerated for three different positions of the doublet, corresponding to the doublet being in the beginning, middle, and end of the beam expansion telescope. The corresponding PFC can be mathematically expressed as a spatio-spectral phase with a functional form $\alpha r^2 (\omega-\omega_0)$. Figure \ref{fig:fig2} (a) shows the measured velocity of the intensity peak along the optical axis for the axiparabola-focused beam, in vacuum, for the $\alpha = -0.0045$ (green, beginning of telescope), $\alpha = 0.0055$ (blue, middle of telescope), and $\alpha = 0.0190$ (orange, end of telescope) cases, in units of \si{fs/mm^2}. Details of the measurement can be found in Ref. \cite{Liberman_OL_2024}. The vacuum velocity of the laser driver does not neatly correspond to the in-plasma velocity of the wakefield, due to the complex geometry of ionization and other nonlinear effects. Yet, as was shown in Ref. \cite{Liberman_NatureCommunications_2025}, a negative PFC corresponds to a wakefield traveling at a higher velocity while a positive PFC corresponds to a slower propagation velocity.

\begin{figure*} [t!]
   \begin{center}
    \includegraphics[width=\linewidth]{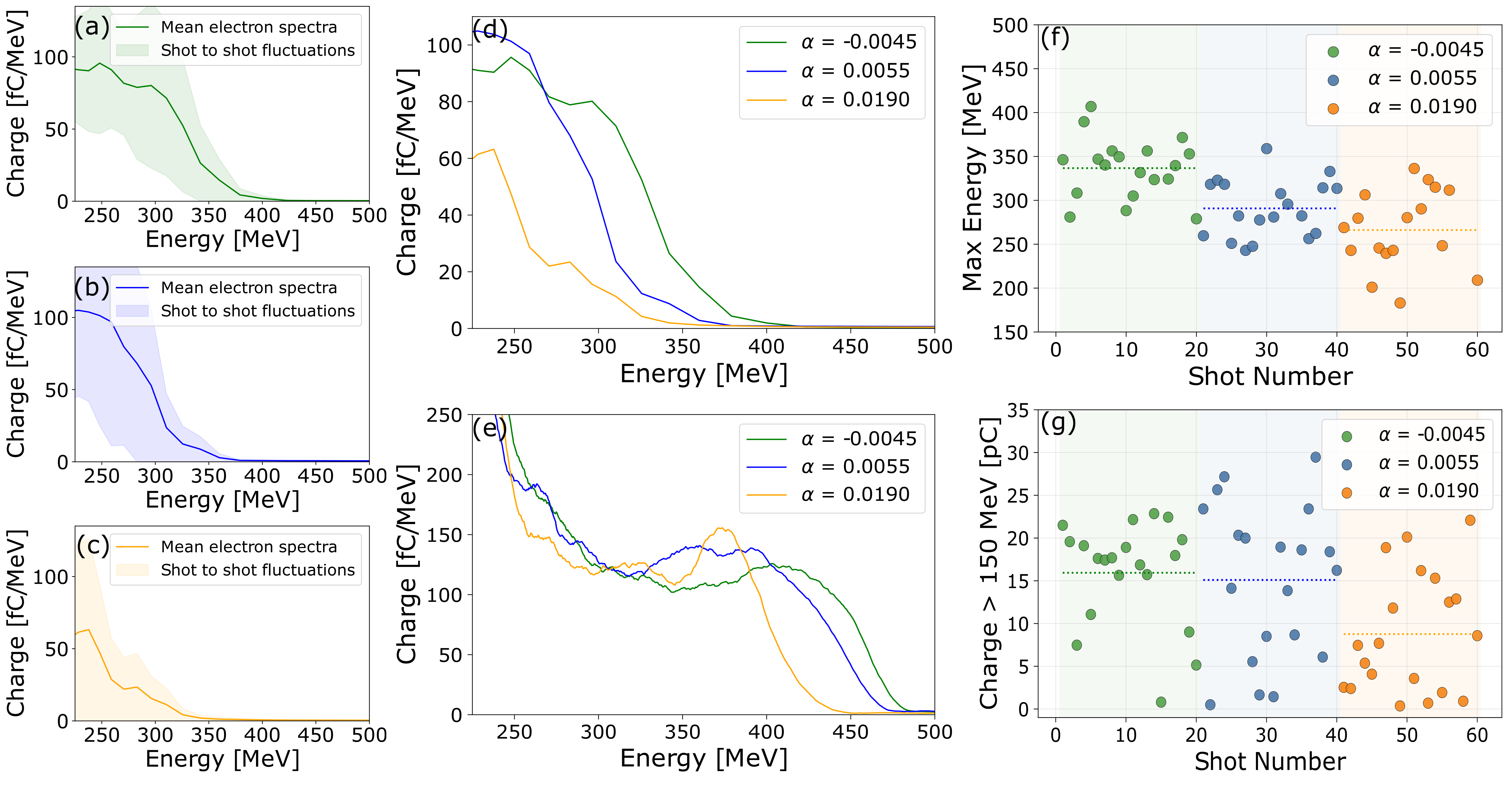}
    \end{center}
    \caption[]{ \label{fig:fig3} (a--c) Electron spectra above $225$ MeV, averaged over 20 shots, for the $\alpha = -0.0045$ (b, green), $\alpha = 0.0055$ (c, blue), and $\alpha = 0.0190$ (d, orange) cases. The energy is shown on the x-axis and the charge density on the y-axis. The shaded area is the RMS shot-to-shot fluctuation. (d) Comparison of the three averaged spectra for the three cases. (e) Comparison of the spectra obtained from the PIC simulation for the three cases. (f) Plot of the maximum energy fluctuations for each of the three cases. Dotted horizontal line gives average maximum energy for each of the cases. (g) Plot of the charge (above $150$ MeV) fluctuations for each of the three cases. Dotted horizontal line gives average charge for each of the cases. }
\end{figure*}

Figure \ref{fig:fig2} (b), (c), and (d), show three selected Lanex scintillator images each for the three different wake velocities. The Lanex scintillator image provides charge, energy, and divergence information for the accelerated electrons. Energy resolution is provided along the x-axis, with the Lanex images in the figure cropped to emphasize the energies between $150-500$ MeV. The y-axis gives information about the divergence of the electron beam. The colorbar corresponds to the charge density. Background subtraction was done for each of the shots. Other than this, they are shown as measured in the experimental setup. The shots were selected among the highest energy of the 20 shots taken for each case. The total charge above $150$ MeV is shown on each Lanex image. As can be seen, for all the cases energies up to several hundred MeV are achieved, with charges in the low tens of picocoulombs.   

A comparison of the $\alpha = -0.0045$ case in figure \ref{fig:fig2} (b) with the $\alpha = 0.0055$ case in figure \ref{fig:fig2} (c) shows that the faster wakefield velocity is able to achieve a higher maximum electron energy. This is further emphasized when comparing with the $\alpha = 0.0190$ case in figure \ref{fig:fig2} (d). Whereas the faster wakefield is able to accelerate electrons reaching up to $400$ MeV, the electrons accelerated by the slower wakefield barely cross $350$ MeV. These Lanex images therefore demonstrate a visible impact that the wake velocity has on the maximum achievable electron energy, pointing towards a partial mitigation of dephasing effects by the faster wakefield.

Figure \ref{fig:fig2} (e) shows PIC simulated Lanex images for the three cases. The simulated Lanex images are quite similar to their experimental counterparts. Critically, the simulated Lanex images show the same dependence of the maximum energy on the wakefield velocity, significantly reinforcing the veracity of the experimental results. The higher cutoff energies seen in the simulation--with a cutoff near $485$ MeV for the $\alpha = -0.0045$ and $440$ MeV for the $\alpha = 0.0190$ case--are likely due to the more idealized gas profile and aberration free axiparabola used in the simulation. The charge is also higher in the simulation, likely for the same reasons. Despite these minor differences, the experiment and the simulation capture the same essential physics, demonstrating the impact of wakefield velocity on the electron acceleration process. 

Figure \ref{fig:fig2} (f) shows a simulated wakefield for the $\alpha = -0.0045$ (top) and $\alpha = 0.0190$ (bottom) cases. The blue colorbar shows the relative electron density distribution $n_e/n_0$, while the red color shows the intensity of the axiparabola-focused laser field. The line plots show the longitudinal acceleration field $E_z$ for the $\alpha = -0.0045$ (green) and $\alpha = 0.0190$ (orange) cases. As can be seen, the wakefields propagate at different velocities, with the $\alpha = -0.0045$ wakefield over a micron further in front than the $\alpha = 0.0190$ wakefield. Importantly, the electrons in the $\alpha = 0.0190$ case are a bit closer to the point where the accelerating field goes to zero than in the $\alpha = -0.0045$ case. Thus, this snapshot provides direct evidence both of the different propagation velocities for different PFCs and of the partial mitigation of dephasing. 

The impact of PFC, and thus of wakefield velocity, on the electron energy is further emphasized when looking statistically at the data. Figure \ref{fig:fig3} (a) shows a graph of the electron energy (plotted above $225$ MeV to emphasize the cutoff energy) versus the charge density for 20 shots, for the $\alpha = -0.0045$ case. The solid line represents the mean charge at each energy while the shaded area gives the RMS of the shot-to-shot fluctuations. Figure \ref{fig:fig3} (b) and (c) show such electron spectra for the $\alpha = 0.0055$ and $\alpha = 0.0190$ cases, respectively. Figure \ref{fig:fig3} (d) show the three mean spectra overlapped. As can be clearly seen, the cutoff energies show a dependence on the wakefield velocity. Over the course of the acceleration, there is around a $50$ MeV difference between the cutoff energies of the slowest wakefield ($\alpha = 0.0190$, orange), at around $350$ MeV, and the fastest ($\alpha = -0.0045$, green solid), at around $400$ MeV. 

Figure \ref{fig:fig3} (e) shows simulated spectra for the three wakefield velocities. As can be seen, the simulated spectra correspond quite well to the averaged experimental spectra in figure \ref{fig:fig3} (d). As with the Lanex images, the spectra show a dependence of maximum energy with wakefield velocity, with a difference of around $45$ MeV between the $\alpha = -0.0045$ and $\alpha = 0.0190$ cases. The close correspondence between the experimental and simulated results and the similar predictions of the dependence of maximum energy with wakefield velocity reinforce the fact that the experiment and simulation capture the same physical processes. 

Figure \ref{fig:fig3} (f) shows the maximum energy for each of the shots shown in figured \ref{fig:fig3} (a--c), thus giving the maximum energy distribution. The dotted horizontal lines give the average maximum energy at each wake velocity, clearly showing the difference between each velocity. Statistically, the distributions of the maximum energies differ in a significant way, with a p-value of $0.0004$ and, thus, a significance of above $3.5\sigma$ between the $\alpha = -0.0045$ and the $\alpha = 0.0055$ cases and even higher when considering the $\alpha = 0.0190$ case. 

Figure \ref{fig:fig3} (g) shows the charge above \SI{150}{MeV} for each shot, with the dotted horizontal lines showing the average for each wakefield velocity. As can be seen, the charge fluctuates significantly for all wake velocities. The energy shift is consistent, despite the wide range of charge fluctuations. This suggests that the effect can't be attributed to beam loading effects or a potential change in the place of injection, but rather to the impact of the wakefield propagation velocity. 

In addition to the comparison with PIC simulations, a simple analytical model was developed to see whether the electron energy dependence on wakefield velocity seen in the data was consistent with theoretical expectations. Further details of the model can be found in Appendix C. While this model can't accurately predict the final electron energy due to the fact that it ignores aspects like the impact of beam loading \cite{Rechatin_PRL_2009} and the complex evolution of the laser pulse \cite{Miller_ScientificReports_2023,Liberman_NatureCommunications_2025}, which can significantly impact the accelerating field, it captures some of the essential physics. Thus, while the model overestimates the maximum energy, it also predicts that the faster propagating wakefield of $\alpha = -0.0045$ will have a higher maximum energy than in the slower propagating wakefields. When conditions similar to the experimental conditions are put into the model, the maximum electron energy in the $\alpha = -0.0045$ case is around $70$ MeV higher than in the $\alpha = 0.0190$ case. 

The successful demonstration of electron acceleration using a flying-focus wakefield represents an important milestone on the road to dephasingless acceleration. It shows that the wakefield generated by an axiparabola-focused beam can maintain the coherent structures necessary for relativistic electron acceleration. Combining this axiparabola-focused beam with spatio-temporal couplings enabled the manipulation of the wakefield propagation velocity. The observation that the maximum electron energy increases with higher wakefield velocities reinforces interest in the flying-focus wakefield to mitigate dephasing effects. The results are further reinforced with particle-in-cell simulations that closely match the experimental predictions and with an analytical model. 

Fully realizing the potential of dephasingless acceleration, however, will require further overcoming some challenges. As stated above, the conversion between the measured vacuum velocity of the intensity peak and the propagation of the wakefield in plasma is highly non-trivial. This means that precisely tuning the wakefield propagation velocity requires fine control over the spatio-temporal couplings and, ideally, the ability to measure the velocity \textit{in situ}. Given the small magnitude of the effect, this remains an experimental barrier to overcome. Furthermore, maintaining a stable wakefield and injecting charge successfully is a more significant challenge in flying-focus pulses, one that still needs to be perfected.

The emergence of a new generation of single-shot spatio-temporal pulse measurements \cite{Smartsev_OL_2024,Howard_NaturePhotonics_2025} will allow for more careful measurement and control of the wakefield velocity. Optimizing this velocity could enable the achievement of multifold gains in the maximum achievable electron energy in LWFA \cite{Palastro_PRL_2020,Caizergues_NaturePhotonics_2020}. In addition to the milestone for dephasingless acceleration, the demonstration of a stable, flying-focus wakefield could serve as inspiration for the exploration of other exotic wakefield configurations, such as the use of helical beams to generate LWFAs for positron acceleration \cite{Vieira_PRL_2014}.

\section*{Acknowledgments}
The authors would like to thank Dr. Igor Andriyash for developing the axiprop code; Dr. Eitan Levine, Dr. Yinren Shou, Ivan Kargapolov and Salome Benracassa for constructive discussions; and Dr. Eyal Kroupp for helping manufacture the setup.

\section*{Data Availability Statement}
The raw data used in this work, as well as the the code, can be made available upon reasonable request to the authors. 

\section*{Conflict of Interest}
The authors declare no conflicts of interest.

\section*{Funding}
The research was supported by the Schwartz/Reisman Center for Intense Laser Physics, the Benoziyo Endowment Fund for the Advancement of Science, the Israel Science Foundation, Minerva, Wolfson Foundation, the Schilling Foundation, R. Lapon, Dita and Yehuda Bronicki, and the Helmholtz Association.

\appendix

\section{Appendix A: Additional Experimental Details}

The gas flow from the slit nozzle was simulated using the Ansys Fluent software, computationally simulating the 3D geometry of the nozzle. A mesh of approximately $10$ million elements was used to solve the flow equations, considering the specific heat, thermal conductivity and viscosity parameters of the gas. The initial pressure at the outlet was $1$ Pa. The pressure at the inlet was varied to correspond to the experimental conditions. The simulation used a pressure of $30$ bars of pure Helium at the inlet. The results provided the density of molecules over the volume of interest that was used as input for the PIC simulations.

Interferometry measurements were also been performed for the slit nozzle, using Argon gas. In order to validate the simulated flow, simulations were run using Argon and the densities obtained were compared. The close correspondence between measurements and simulations confirmed the accuracy of the Ansys Fluent simulation.

After exiting the gas jet, the electron bunch was passed through a 20-cm-long 1 T magnetic dipole in order to induce an energy dependent angle in the electron bunch. After further free-space propagation, the electrons spatially spread out along the x-axis according to their energies. The electrons then impinged upon a Lanex scintillating screen which was imaged by a Hamamatsu ORCA-FLASH4.9 digital CMOS camera. Through simulations based on a measured map of the magnetic field in the magnet, the $x$-axis position of the electrons on the screen was calibrated to specific electron energies. The $y$-axis spread yielded information about the divergence of the electron bunch. The Lanex screen was calibrated with a radioactive tritium source, allowing the conversion from pixel brightness to electron bunch charge. Since the path length between the source and the Lanex is not constant over the different energies, the stated divergence axis is for the center-point of the Lanex. At either extreme of the Lanex, this would introduce an error of around 10\% to the stated divergence values. RMS pointing fluctuation was around 5.5 mrad and the divergence was around 1.5 mrad. The spectra shown are binned as to account for the energy uncertainly in the lanex. The uncertainty of the cutoff energy of the electrons is around 12 MeV, significantly less than the observed energy splitting.

As discussed above, the PFC was controlled via a specialized doublet lens, inserted into the final beam expansion telescope of the laser system. Further details of the PFC control system can be found in Ref. \cite{Smartsev_JoO_2022}. Moving the doublet introduced a slight change to the focusing term, resulting in a \SI{400}{\um} focal shift between the most negative and most positive PFCs. This focal shift was measured and is predicted by Ansys Zemax OpticStudio simulations. To compensate for this focal shift, the gas jet was moved the same amount in the experiment, in order to have the laser focus at the same position inside of the gas jet. In addition, to ensure that PFT does not play a role in the acceleration, PFT was optimized at each of the doublet positions. Focal spot scans were performed at each of the PFC values, generating a 3D fluence map of the laser at focus. No significant difference was found between the pulses for the different PFC values, indicating that changing the PFC value didn't lead to a change in the transverse fluence, intensity distribution, and mode structure.

\section{Appendix B: Additional Simulation Details}

The axiparabola-focused beam was initialized by using the Axiprop code \cite{Andriyash_Axiprop,Oubrerie_JoO_2022} to propagate a simulated pulse from the axiparabola surface to the focal plan. The PFC was accounted for by adding a phase correction to the laser before the reflection. To accommodate Axiprop's axisymmetric solver and the angular mode decomposition used in the quasi-3D spectral code FBPIC \cite{Lehe_ComPhysCom_2016}, an on-axis axiparabola with corresponding parameters to the experiment (\SI{480}{\mm} focal length, \SI{5}{\mm} focal depth for \SI{25}{\mm} radius) was simulated. The laser profile of the initial pulse was a 16th order super-Gaussian in the transverse plane and the pulse energy was \SI{1.5}{J}. The temporal profile was a \SI{27}{\fs} FWHM Gaussian and the central wavelength was \SI{800}{\nm}. 

The output of the axiprop simulation was saved in LASY format \cite{thevenet_lasy} and was then used as the input into FBPIC simulations \cite{Lehe_ComPhysCom_2016}, assuming linear polarization in the $x$ direction. In order to ensure that the laser was not artificially cut in the simulation, a large simulation box of \SI{558}{\um} in the transverse $r$ direction and \SI{150}{\um} in the longitudinal $z$ direction was used, in addition to the use of two azimuthal modes. The resolution of the grid was $dz = \SI{0.04}{\um}$ and $dr = \SI{0.24}{\um}$. In order to decrease computational time, a Lorentz-boosted frame was employed, with a boost factor $\gamma = 3$ \cite{Lehe_2016_PRE_94_53305}.

The gas density profile was taken from the output of the Ansys Fluent fluid simulation, shown in figure \ref{fig:setup} (d), with a peak background electron density of \SI{4.2e18}{cm^{-3}}. A gas mixture of helium and nitrogen was used, with the macroparticle weights arranged to give a molecular share of 3\% of $N_2$. The gas was initialized to be neutral, in order to account for the impact of diffraction from a partially ionized gas at larger radii. The gas was initialized with 32 helium and 32 nitrogen neutral atomic particles per 2D cell (2 in $r$, 2 in $z$, 8 in $\theta$ directions, respectively) within the radius of \SI{30}{\um} and 8 helium and 8 nitrogen atomic particles (1 in $r$, 1 in $z$, 8 in $\theta$) outside this radius. The axiparabola was initialized to focus at a starting depth of \SI{4}{\mm} into the simulated profile. 

To simulate the Lanex image, the electrons generated in the PIC simulation were propagated through a model of the spectrometer magnet and the resultant charge density was projected onto a planar screen. The parameters of the magnet and the screen corresponded to the experimental setup, thus allowing the spectrometer calibrations to be matched.

\section{Appendix C: Analytical Model}

\counterwithin{equation}{section}
\renewcommand\theequation{\thesection\arabic{equation}}
According to Ref. \cite{Lu_PhysRevSTAB_2007}, in the bubble regime we can generally assume a linear relationship between the $E$-field and the coordinate $\zeta = z-z_0(t)$ is assumed, where $z_0$ is the coordinate of the center of the bubble. The $E$-field scales as:
\begin{equation}
    E_z(\zeta) =  \frac{m\omega_p^2}{2e}\zeta = \frac{e n_e}{2\epsilon_0} \zeta, \tag{C1}
\end{equation} 
where $\epsilon_0$, $m$, $e$, and $\omega_p$ are the vacuum permittivity, the electron mass, the electron charge, and the plasma frequency, respectively.

From Ref. \cite{Esarey_ReviewOfModernPhysics_2009}, the group velocity of the laser driver inside of the plasma: 
\begin{equation}
\label{eq:velocity2}
    \frac{v_\mathrm{gr}}{c} = \bigg[1-\frac{n_e}{n_c}\bigg]^{0.5} = \bigg[1-\frac{\omega_p^2}{\omega_0^2}\bigg]^{0.5},\tag{C2}
\end{equation}where $\omega_0$ is the laser frequency. For a density of \SI{4e18}{cm^{-3}}, this would give a velocity of around $0.9989c$. We assume the accelerated electrons are ultrarelativistic ($dz/dt \approx c$) compared to the wake velocity. We also assume that the phase velocity of the wake is equal to the sum of the regular group velocity and the axiparabola-beam correction $\Delta v(z)$ (calculated in vacuum and depending on the PFC value $\alpha$ \cite{Liberman_OL_2024,Oubrerie_JoO_2022}), as demonstrated in Fig.~\ref{fig:fig2} (a), 
\begin{equation}
    v_\mathrm{ph}(z) = v_\mathrm{gr} + \Delta v(z).\tag{C3}
\end{equation}
Solving for $\zeta$ as a function of time yields: \begin{align}
    &\frac{d\zeta}{dt} = c-v_\mathrm{ph}(ct), \tag{C4} \\
    &\zeta(t) = \zeta_\mathrm{min} + \int\limits_{0}^{ct} \left[1-\frac{v_\mathrm{ph}(z)}{c}\right] dz.  \tag{C5}
\end{align}
where $\zeta_\mathrm{min} < 0$ is the $\zeta$ at the point of the electron injection. Now the momentum gained in the longitudinal direction can be solved for:
\begin{align}
&\frac{dp_z}{dt} = -eE_z \implies \frac{dp_z}{dz} = -\frac{e}{c}E_z = -\frac{e^2 n_e}{2\epsilon_0 c}\zeta, \tag{C6}\\
&p_z = -\frac{e^2 n_e}{2\epsilon_0 c}\int\limits_{0}^{z}\bigg(  \zeta_\mathrm{min} + \int\limits_{0}^{z'} \left[ 1-\frac{v_\mathrm{ph}(z'')}{c}\right] dz''\bigg)dz' .\tag{C7}
\end{align}
As stated above, to get the phase velocity of the wake in plasma, the measured velocity profiles in vacuum, shown in figure \ref{fig:fig2} (a), are modified. The assumption is made that the in-plasma profiles correspond to the vacuum profiles minus the difference between the velocity of the laser driver in vacuum and plasma, as obtained in equation \ref{eq:velocity2}. While this assumption ignores some of the impact of the plasma on the velocity, it gives a general correspondence to the wakefield velocity.

\section*{References}

\bibliographystyle{unsrt}
\bibliography{Refs}

\end{document}